\documentclass[a4paper,11pt]{article}
\usepackage{pos}
\usepackage{wrapfig}
\usepackage{xspace}

\def\q{$\bullet$~}
\def\pao{Pierre Auger Observatory\xspace}
\def\Xmax{X_{\rm max}}
\def\gcm2{g/cm$^2$}
\def\Caption#1{\caption{\small #1}}
\def\fref#1{Fig.\,\ref{#1}}

\def\aaa#1#2#3#4#5#6#7{{\sc\scriptsize #1}, {\scriptsize #2, #3
\ifx#4\empty
\else 
~\& #4, #5
\fi 
\ifx#6\empty
\else 
~\& #6, #7
\fi 
.}\par
}

\title{GCOS - The Global Cosmic Ray Observatory}
\ShortTitle{GCOS - The Global Cosmic Ray Observatory}

\author*[a,b,c]{J\"org R. H\"orandel}
\affiliation[a]{Department of Astrophysics/IMAPP, Radboud University, P.O. Box 9010, 6500 GL Nijmegen, The Netherlands}
\affiliation[b]{Vrije Universiteit Brussel, Brussel, Belgium}
\affiliation[c]{Nikhef, Amsterdam, The Netherlands}

\forColl{GCOS} 

\emailAdd{jorg.horandel@ru.nl}

\abstract{
Nature is providing particles with energies exceeding 100 EeV. Their existence
imposes immediate questions: Are they ordinary particles, accelerated in
extreme astrophysical environments, or are they annihilation or decay products
of super-heavy dark matter or other exotic objects?  If the particles are
accelerated in extreme astrophysical environments, are their sources related to
those of high-energy neutrinos, gamma rays, and/or gravitational waves, such as
the recently observed mergers of compact objects? The particles can also be
used to study physics processes at extreme energies; is Lorentz invariance
still valid? Are the particles interacting according to the Standard Model or
are there new physics processes? The particles can be used to study hadronic
interactions (QCD) in the kinematic forward direction; what is the cross
section of protons at center-of-mass energies $\sqrt{s} > 100$~TeV?

These questions are addressed at present by installations like the Telescope
Array and the \pao. After the year 2030, a next-generation
observatory will be needed to study the physics and properties of the
highest-energy particles in Nature, building on the knowledge harvested from
the existing observatories. It should have an aperture at least an order of
magnitude bigger than the existing observatories. 

Recently, more than 200 scientists from around the world came together to discuss the future of the field of multi-messenger astroparticle physics beyond the year 2030. 
Ideas have been discussed towards the physics case and possible scenarios for detection concepts of the
Global Cosmic Ray Observatory - GCOS.
A synopsis of the key results discussed during the brainstorming workshop will be presented.
}

\FullConference{37$^{\rm{th}}$ International Cosmic Ray Conference (ICRC 2021)\\
		July 12th -- 23rd, 2021\\
		Online -- Berlin, Germany}

\begin{document}
\maketitle

\section{Introduction}
Nature is providing particles at enormous energies, exceeding $10^{20}$~eV -- orders of magnitude beyond the capabilities of human-made facilities like the Large Hadron Collider (LHC at CERN).
At the highest energies the precise particle types are not yet known, they might be ionized atomic nuclei or even neutrinos or photons. Even for heavy nuclei (like, e.g., iron nuclei) their Lorentz factors $\gamma=E_{\rm tot}/m c^2$ exceed values of $10^9$.
The existence of such particles imposes immediate questions, yet to be answered:
\q What are the physics processes involved to produce these particles?
\q Are they decay or annihilation products of dark matter?
\cite{Alcantara:2019sco,Aloisio:2015lva}
If they are accelerated in violent astrophysical environments: \q How is Nature being able to accelerate particles to such energies?
\q What are the sources of the particles? Do we understand the physics of the sources?
\q Is the origin of those particles connected to the recently observed mergers of compact objects -- the gravitational wave sources? \cite{LIGOScientific:2017ync,Lipari:2017qpu,Branchesi:2016vef,Gergely:2007ny,Gergely:2008dw,Gergely:2010xr,Tapai:2013jza}
The highly relativistic particles also provide the unique possibility to study (particle) physics at its extremes:
\q Is Lorentz invariance (still) valid under such conditions?
\cite{Klinkhamer:2008ss,Klinkhamer:2017puj,Aloisio:2000cm,Cowsik:2012qm,Martinez-Huerta:2016azo,pao-liv}
\q How do these particles interact?
\q Are their interactions described by the Standard Model of particle physics?
When the energetic particles interact with the atmosphere of the Earth, hadronic interactions can be studied in the extreme kinematic forward region (with pseudorapidities $\eta>15$):
\q What is the proton interaction cross section at such energies ($\sqrt{s}>10^5$~GeV)?

The highly energetic particles, called ultra high-energy cosmic rays (UHECRs), are extremely rare: their flux is lower than one particle per square kilometer per century above $7\times10^{19}$~eV \cite{Bluemer:2009zf,Anchordoqui:2018qom,AlvesBatista:2019tlv}. To study their properties, large detection facilities are needed in order to collect a reasonable number of them in an acceptable time span.
At present, the largest detector is the \pao in Malarg\"ue, Argentina \cite{ThePierreAuger:2015rma}, covering an area of 3000 km$^2$.
To increase its sensitivity to the type of particle, at present, additional components are being installed at the Observatory \cite{Aab:2016vlz,Castellina:2019irv,Horandel:2019hrm}.
In the Northern Hemisphere the Telescope Array \cite{Bergman:2020izr}, located in Utah, USA, is covering an area of 700 km$^2$, presently undergoing an extension \cite{Kido:2019enj} to cover about 2800 km$^2$.
The goal of these installations is to measure the properties of UHECRs with unprecedented precision in the next decade. Key properties include the arrival direction (on the sky), the energy, and the particle type. When the ultra-high-energy (UHE) particles enter the atmosphere of the Earth they undergo (nuclear) interactions and produce avalanches of secondary particles, the extensive air showers. Secondary products of these air showers are measured with ground-based detectors. This makes it demanding to determine the particle properties, in particular, identifying the particle type is an experimental challenge. It requires an elaborate concept to simultaneously measure several components of the air showers \cite{Aab:2016vlz,Castellina:2019irv,Horandel:2019hrm}.
With the existing (upgraded) facilities it is expected to measure a few dozens of particles at the highest energies ($>10^{20}$~eV) and identify their type. This work is expected to continue until $\sim 2030$.

In May 2021 more than 200 scientists came together for an online workshop to discuss the prospects for
multi-messenger astroparticle physics at ultra-high energies beyond the year 2030 \cite{gcos2021}.
In the following we summarize the key findings of this brainstroming workshop, defining a possible path towards a Global Cosmic Ray Observatory -- GCOS.\footnote{Detailed information and more references are available online \cite{gcos2021}, only a brief glance can be given here.}

\section{Towards a science case for ultra-high-energy particles}
We are living in a golden epoch in Astrophysics where we have witnessed the birth and the first steps in the development of multi-messenger astronomy \cite{gcos2021-lipari}.
Our understanding of the high-energy Universe has significantly expanded and progressed thanks to observations obtained recently with different messengers in a broad range of energies.
There are several strong motivations for the proposal of new high-sensitivity observation programs with different messengers in different energy ranges.
The corresponding discussion is timely and necessary but far from easy as different people have different motivations and priorities for the study of the high-energy Universe.
This diversity of perspectives should be considered as a gain for the field.
New territory is being explored where new, also surprising, phenomena -- e.g. dark matter, exotic objects, violations of symmetries such as Lorentz invariance are possible.
Even in the most conservative case the high-energy sources are amazing objects that challenge our view and constitute unique laboratories to test the fundamental laws of physics.
A program for future observations should be constructed to contribute significantly to the understanding of high-energy sources, which is already of great interest in the most conservative case, let alone the case of exotic phenomena of new physics, which we anticipate as an obviously exciting additional possibility.

\begin{wrapfigure}{r}{0.5\textwidth} 
  \vspace{-6mm}
  \includegraphics[width=0.5\textwidth]{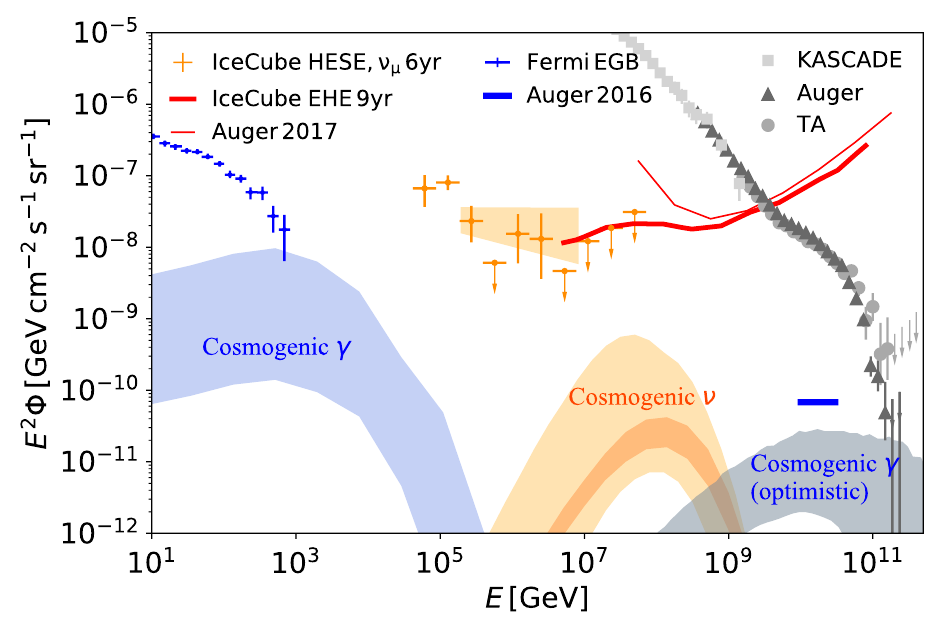}
  \vspace{-5mm}
  \Caption{Energy flux per steradian in the local Universe contained in various messengers \cite{AlvesBatista:2019tlv}.}
  \vspace{-4mm}
  \label{fig:edensity}
\end{wrapfigure}
The following questions need to be answered:
Where are we now and where will we be in 2030 with our understanding of the high-energy Universe?
There has been enormous progress, but many essential questions remain to be answered.
What are the contributions of the observations at the highest energies to multi-messenger astrophysics?
And how do they fit in the general picture?
What are the best directions to make progress?

Three messengers are ``inextricably'' tied together (cosmic rays, gamma rays, high-energy neutrinos) and provide complementary information about the same underlying physical phenomena. This is also visible when the energy density in various messengers is considered, as illustrated in \fref{fig:edensity}.
Transient sources associated with the formation of compact objects provide a link to a fourth messenger: gravitational waves.
Electrically charged particles can acquire very large energies propagating in the electromagnetic fields of astrophysical objects/transients.
Neutrinos and gamma rays are generated with approximately equal rates in the decay of pions and other particles, created in the interactions of protons and nuclei (hadronic mechanism).
Gamma rays are also created by radiation processes of relativistic electrons/positrons (leptonic mechanism).
Thus, gamma rays and neutrinos trace the populations of relativistic charged particles such as protons, nuclei, electrons, and positrons in the sources.
The relation of the fluxes of neutrino and gamma rays reflects the relative importance of the acceleration of electrons/positrons versus protons/nuclei and, in addition, effects like absorption of photons inside the sources and during propagation, neutrino oscillations, and possibly other physical phenomena.
The relation between the fluxes of cosmic rays observed at the Earth and the gamma-ray and neutrino fluxes is a much more difficult and less understood problem because it depends on the escape of cosmic rays from the sources and propagation in the Milky Way or/and extra-galactic space with large uncertainties for both environments.
In the coming decade(s) it will also be of importance to link the recent detailed observations of black holes \cite{Akiyama:2019cqa,Akiyama:2021qum} to the formation of jets from AGNs and their multi-wavelength observations \cite{Acciari:2009rs} and, thus, to the acceleration of UHECRs in such objects.
It will also be important to better understand gravitational wave sources and their possible role as sources of the highest energy particles in the Universe.

To formulate a detailed science case, a series of points needs to be addressed as outlined in the following.

\subsection{Theoretical requirements}
\paragraph{Understanding the effects of Galactic and extra-galactic magnetic fields}
A key component for a science case will be to be able to backtrack charged cosmic rays in the Galactic and extra-galactic magnetic fields. This requires detailed knowledge of the structure of cosmic magnetic fields. 
In the next decade we can expect an improvement of Galactic magnetic field models \cite{gcos2021-haverkorn}. For small-scale magnetic fields correlations between magnetic field orientations as measured with different tracers and in different media are important. Modeling is slowly progressing beyond Gaussian random fields. In the future it is expected that models include ``proper'' turbulent fields and their correlations to cosmic rays, thermal electrons, etc.
For large-scale magnetic fields, the IMAGINE consortium will model Galactic magnetic fields and compare models quantitatively \cite{Boulanger:2018zrk}. IMAGINE aims to produce the best-fit large-scale Galactic magnetic field model and UHECR arrival directions can be corrected.
Also extra-galactic magnetic fields can significantly influence the backtracking to sources at large distances \cite{AlvesBatista:2017vob}.
Therefore, modelling their seeding through possible primordial fields and through astrophysical processes, as well as their evolution
and resulting structure needs to be improved by detailed numerical simulations of structure formation including magnetohydrodynamics.

\begin{wrapfigure}{r}{0.5\textwidth} 
  \vspace{-2mm}
  \includegraphics[width=0.5\textwidth]{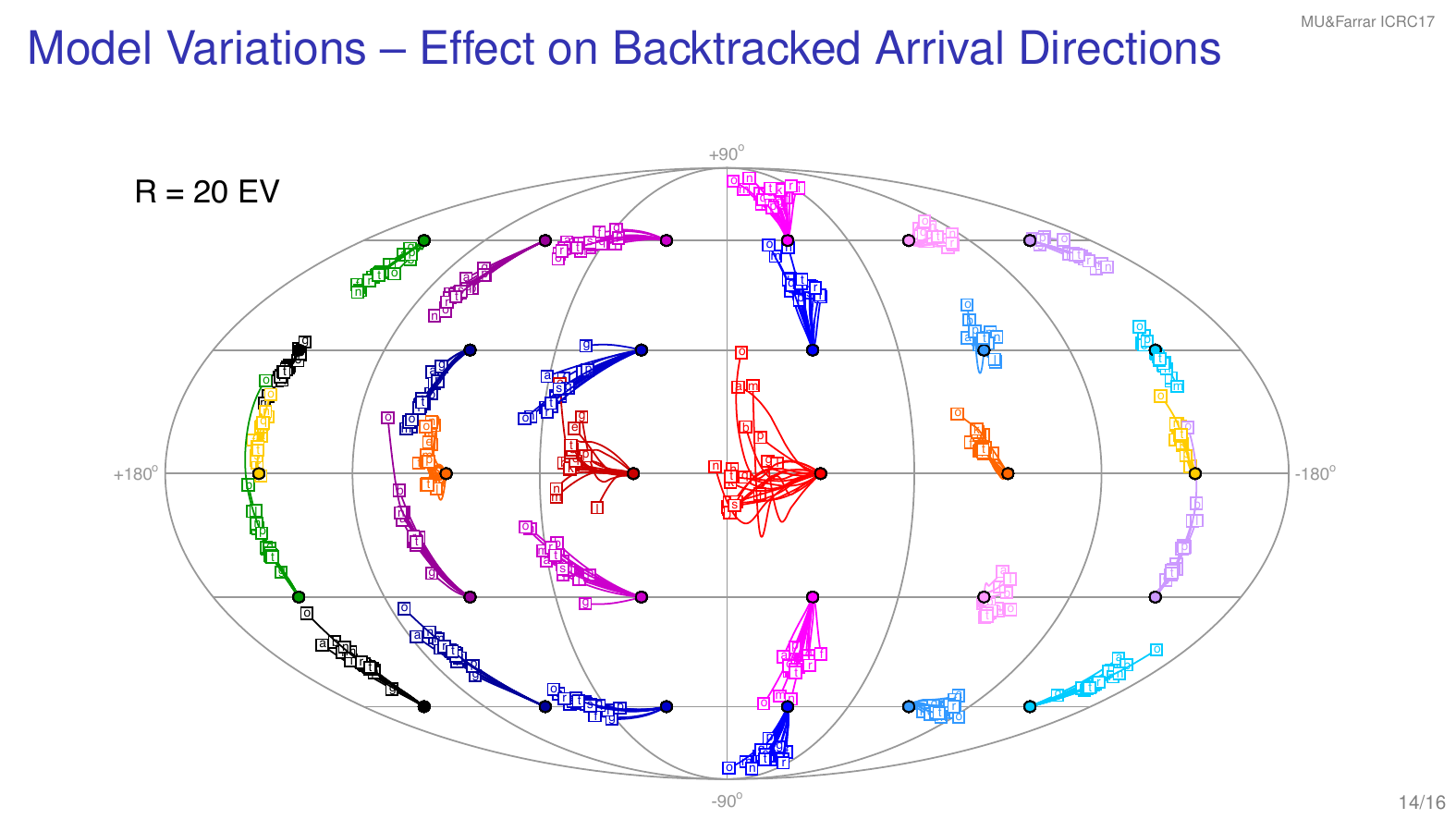}
  \vspace{-5mm}
  \Caption{Backtracking of charged particles with $R=20$~EV from a regular grid of initial directions (dots) through different models of the coherent Galactic magnetic field. See \cite{Unger:2017wyw} for details.}
  \vspace{-3mm}
  \label{fig:bfield}
\end{wrapfigure}
To conduct charged-particle astronomy \cite{gcos2021-unger,gcos2021-farrar} it is also desired to have
a large exposure to reach high rigidity values and the ability to determine the charge for each measured cosmic ray.
GCOS will need good charge calibration. It will be important to reduce the uncertainties introduced by the hadronic interaction models. Also model-independent methods should be developed, which allow a charge calibration based on measured data, such as, e.g., the Gerasimova-Zatsepin effect \cite{Lafebre:2008rz}.
Prospects are good that in the GCOS era it will be possible to track high-rigidity charged particles back to their sources.
This is illustrated in \fref{fig:bfield} for particles with a rigidity of $R=20$~EV in the Galactic magnetic field \cite{Unger:2017wyw}.

Other important questions are on the number of sources, both the expected number at highest energies but also the number required to sustain the observed power in extraglactic cosmic rays.
A simple estimate, taking into account the energy output of possible sources as, e.g., the kinetic energy in AGN jets hints towards a modest number of sources (order of 10 to 25) in order to sustain the power measured in extra-galactic cosmic rays \cite{Hoerandel:2008re}.

\paragraph{Interaction properties}
It will be crucial to improve our understanding of the interactions of particles with extreme energies. This understanding is needed for a better description of the physics of the UHE particle sources, the particle propagation through intergalactic and interstellar space, as well as finally, the particle interactions within the Earth's atmosphere and the development of air showers \cite{gcos2021-photonuc, gcos2021-pierog}.
Codes used to model the UHE particle propagation, such as, e.g., CRPropa \cite{Batista:2016yrx} or SimProp \cite{Aloisio:2017iyh} require detailed knowledge of photo-nuclear processes as well as a good description of the diffuse astrophysical backgrounds over a broad frequency range from the radio regime to gamma-rays. Processes, such as photo-disintegration and photo-meson production need to be known in detail, since the understanding of the UHECR characteristics might be affected by uncertainties in the description of these interactions \cite{Batista:2015mea, Boncioli:2016lkt,PierreAuger:2016use}. To interpret air shower data it is crucial to understand (hadronic) interactions at extreme energies, well beyond LHC energies, and, in particular in the kinematic forward region. Of particular importance for GCOS will be dedicated  efforts at the LHC such as proton-oxygen collisions \cite{Dembinski:2020dam}.
It is also desired to improve the knowledge of source physics to justify assumptions made for the propagation models such as, e.g., the exact shape of the energy spectra in the fall-off region and the question: do the spectra for different nuclei follow a Peters' cycle \cite{peters}?

\paragraph{Multimessenger connections} GCOS will be capable of detecting neutrinos and photons, greatly enriching its science case. In the multimessenger era, it will be an important partner to search for neutral UHE particles associated with transient events such as mergers of compact binaries, tidal disruption events, and gamma-ray bursts, among others, providing insights into the most energetic processes in Nature~\cite{Aab:2019gra}.  In addition, GCOS will either measure or constrain the fluxes of cosmogenic neutrinos and photons, consequently improving our understanding about UHECR sources (see, e.g.,~\cite{Romero-Wolf:2017xqe, AlvesBatista:2018zui, vanVliet:2019nse}).

\paragraph{Model scenarios}	
We will prepare model scenarios and it needs to be demonstrated in a few examples that the UHE particle sources can be found.
We need to prepare examples and show that we can correct for the magnetic fields within the Galaxy and beyond and backtrack the particles to their sources.
Recently, the correlation of the arrival direction of cosmic rays with objects in astronomical catalogues has been investigated \cite{Aab:2018chp} and it has been found that a ``smearing angle'' of about $15^\circ$ is sufficient to find a significant correlation. Thus, if the knowledge about cosmic magnetic fields will allow correction for deflections on the $10^\circ - 20^\circ$ scale, this would dramatically improve the ability to backtrack the particles and conduct particle astronomy.
Such model scenarios will need to consider in detail the physics of various sources, the acceleration mechanisms taking place, the physics which governs the escape of particles from the source region, the particle propagation through intergalactic and interstellar space until their interactions with the atmosphere of the Earth. Ideally, full end-to-end simulations will be prepared for different source classes, such as AGNs, gamma-ray bursts, and gravitational-wave sources. Such simulations will yield quantitative estimates for the quality required of the measurable quantities, such as the angular, energy, and mass resolution of GCOS.

\subsection{What will we learn in the next decade from TAx4 and AugerPrime?}
It needs to be carefully evaluated and extrapolated what we will learn from existing observatories (and their upgrades), in particular from the Telescope Array ($\times4$) and the \pao.
The role of GCOS will depend on the outcome of these facilities in the coming years. If sources will be detected in the next decade, then detailed studies of their characteristics will be possible, and particle astronomy will become feasible. On the other hand, if no sources are found and if the UHECR mass composition at the highest energies is relatively heavy (implying a large $Z$) it will be an enormous challenge to backtrack particles with relatively low rigidities.

To specify a detailed GCOS science case we  need to follow how the observed anisotropies \cite{Aab:2017tyv,Aab:2018chp,Abbasi:2018qlh,Abbasi:2020fxl} develop over time and try to extrapolate this behavior beyond 2030.
It needs to be simulated how a large-area detector such as GCOS would observe these anisotropies under given assumptions \cite{gcos2021-taaniso}.
It needs to be evaluated how long it will take to achieve higher significances for observed structures on the sky such as, e.g., a clustering around Cen A.
To interpret the observed anisotropy on the sky it will be important to clarify if the sources are transient or continuous.
It also needs to be clarified what happens if a correlation would be found: can we make unambiguous claims? 
It should also be noted that the understanding of the extreme Universe at the highest energies necessitates a good understanding of the physics and the underlying processes at the transition from a Galactic to an extra-galactic origin of cosmic rays.

\subsection{Complementary science cases}
A set of  complementary science cases will play an important role for a future UHE particle observatory.

\paragraph{Dark-matter searches} 
For many decades, the favored models characterized dark matter (DM) as a relic density of weakly interacting massive particles (WIMPs). However, LHC experiments have run extensive physics searches for WIMP signals which have returned null results. In addition, a broad WIMP search program has been developed with direct and indirect detection methods, which so far yield to null results. Despite the fact that a complete exploration of the WIMP parameter space remains the highest priority of the DM community, there is also a strong motivation to explore alternatives to the WIMP paradigm \cite{Alcantara:2019sco}. 
Among the well-motivated ideas for what DM could be, the WIMPzilla hypothesis postulates that DM is made of gravitationally produced (nonthermal relic) superweakly interacting supermassive $X$-particles. 
Super-heavy dark matter (SHDM) can be produced from quantum fluctuations at the end of inflation, similarly to the the observed cosmic microwave background fluctuations. SHDM tends to decay with a high photon/hadron ratio and thus predicts a detectable photon flux if its lifetime is sufficiently low.
Experimentally, the key will be to measure the photon/hadron ratio at the highest energies which would require a resolution in $\Xmax$ better than $\Delta\Xmax<30$~\gcm2. As illustration, using data from the \pao, already strong limits could be set on the lifetime of hadronically decaying SHDM particles for the parameter space $10^{14}\le M_X/\mbox{GeV} \le 10^{16}$.
To further constrain SHDM scenarios it will be crucial that GCOS will have photon-detection capabilities to constrain, e.g., the flux of photons from certain regions, such as the Galactic center \cite{Anchordoqui:2021crl, gcos2021-dm}.
Another class of DM candidates are dark nuclearites or dark quark nuggets with signatures detectable by a GCOS-type observatory \cite{Anchordoqui:2021xhu}.


\paragraph{Fundamental physics and quantum gravity}
UHE particles can be used as probes of fundamental physics and quantum gravity. 
The data can be used to search for Lorentz symmetry violations in the nucleon or photon sector \cite{gcos2021-liv}. 
The study of the interactions of UHECRs with universal diffuse background radiation can provide very stringent tests of the validity of special relativity \cite{Aloisio:2000cm}. For example, it can be investigated if the dispersion relation between energy, momentum, and mass are modified by non-renormalizable effects at the Planck scale. 
This would influence the energy at which the GZK effect occurs.
Thus, the experimental confirmation of the existence of structures in the UHECR energy spectrum can, in principle, put very stringent limits on the scale where special relativity and/or continuity of space-time may possibly break down.

Another important aspect are effects of Lorentz symmetry violation on air showers. The main idea is that modified decay rates of neutral and/or charged pions and muons can change the shower characteristics, such as the muon content and $\Xmax$. This could also include threshold effects, as, e.g., in the muon content as a function of primary energy \cite{Klinkhamer:2017puj} (see also following topic).

\paragraph{Particle physics} 
\begin{wrapfigure}{r}{0.5\textwidth} 
  \vspace*{-2mm}
  \includegraphics[width=0.5\textwidth]{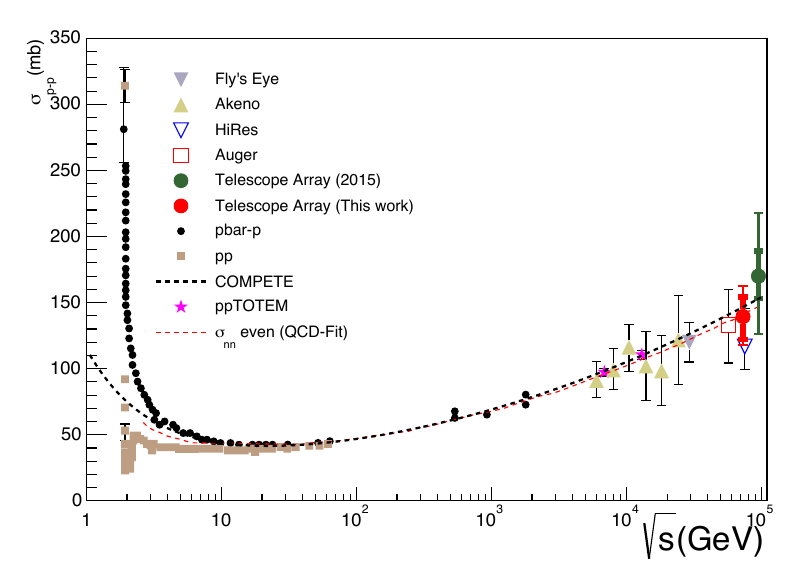}
  \vspace*{-7mm}
  \Caption{Measured values of the proton-proton cross section as a function of the center-of-mass energy from air shower and accelerator measurements, for details see \cite{Abbasi:2020chd}.}
  \vspace{-5mm}
  \label{fig:crosssection}
\end{wrapfigure}
One of the experimental challenges in determining the mass of cosmic rays from air shower measurements is the degeneracy between the mass of the incoming particles and hadronic interactions \cite{gcos2021-pierog}. In principle, they effect the same basic properties like cross sections etc. Inconsistent results on the mass composition of cosmic rays (e.g.\ \cite{Horandel:2003vu,Kampert:2012mx}) point to weakness of models used to describe hadronic interactions. Thus, hybrid measurements of air showers are mandatory for GCOS to verify hadronic interaction models \cite{Cazon:2020zhx}.

For example, the hybrid design of the \pao allows for detailed measurements of the air shower properties and, thus, investigations of the accuracy of hadronic interaction models. The muon content of air showers and its fluctuations have been analyzed in detail \cite{PierreAuger:2014zay,PierreAuger:2014ucz,PierreAuger:2016nfk,PierreAuger:2017tlx,Aab:2020frk,PierreAuger:2021qsd}, indicating a deficit of muons in the predictions of post-LHC hadronic interaction models.

Air shower data are also used to measure the cross sections for proton-air and proton-proton collisions at center-of-mass energies far above values reachable at accelerators, see \fref{fig:crosssection} \cite{flyseyewq84,Collaboration:2012wt,Abbasi:2020chd}.
To test hadronic interactions at the highest energies with GCOS it will be necessary to have good measurements of both, the electromagnetic and muonic shower components and their arrival times.
The uncertainties in $\Xmax$ are mainly due to the extrapolations of the properties of nuclear collisions. To improve the situation we need precise measurements of the inelastic cross section, multiplicity, and diffraction in $pA$ and $AA$ collissions with $A<20$. Dedicated measurements at the LHC and future colliders will be needed.


\paragraph{Geophysics and atmospheric science}
GCOS will also be able to address scientific questions from the areas of geophysics and atmospheric science. 
An example is the study of ELVES which are a class of transient luminous events, with a radial extent typically greater than 250~km, that occur in the lower ionosphere above strong electrical storms. They have been investigated with the fluorescence detector of the \pao \cite{Aab:2020nkb}.

Another example are high-resolution observations of downward-directed terrestrial gamma-ray flashes (TGFs) detected by the Telescope Array, obtained in conjunction with broadband VHF interferometer and fast electric field change measurements of the parent discharge \cite{Abu-Zayyad:2020abs}. The results show that the TGFs occur during strong initial breakdown pulses (IBPs) in the first few milliseconds of negative cloud-to-ground and low-altitude intracloud flashes, and that the IBPs are produced by a newly-identified streamer-based discharge process called fast negative breakdown. 

The surface detector of the Pierre Auger Observatory has collected some very peculiar events \cite{Colalillo:2017lnj}. 
The signals produced by these events in the detector stations are very long-lasting compared to those produced by cosmic rays. For many events, the number of stations with long signal is big, and they are arranged in a circular shape with a radius of about 6~km. Moreover, a correlation with lightning was observed.

Radio antennas allow detailed insights into the spatial and time structure of the development of lightning strokes in the Earth's atmosphere \cite{lofar-lightning}.

\subsection{Experimental requirements}

To prepare a concrete concept for a design and layout of a UHE particle observatory requires to define a set of requirements which are needed to obtain the objectives identified in the physics case.
What are the minimum requirements to find and study the sources of UHE particles?
Different detection concepts are at hand and their strength and  weaknesses need to be carefully evaluated with respect to the science goals.

\begin{wrapfigure}{r}{0.5\textwidth} 
  \vspace{-2mm} 
  \includegraphics[width=0.5\textwidth]{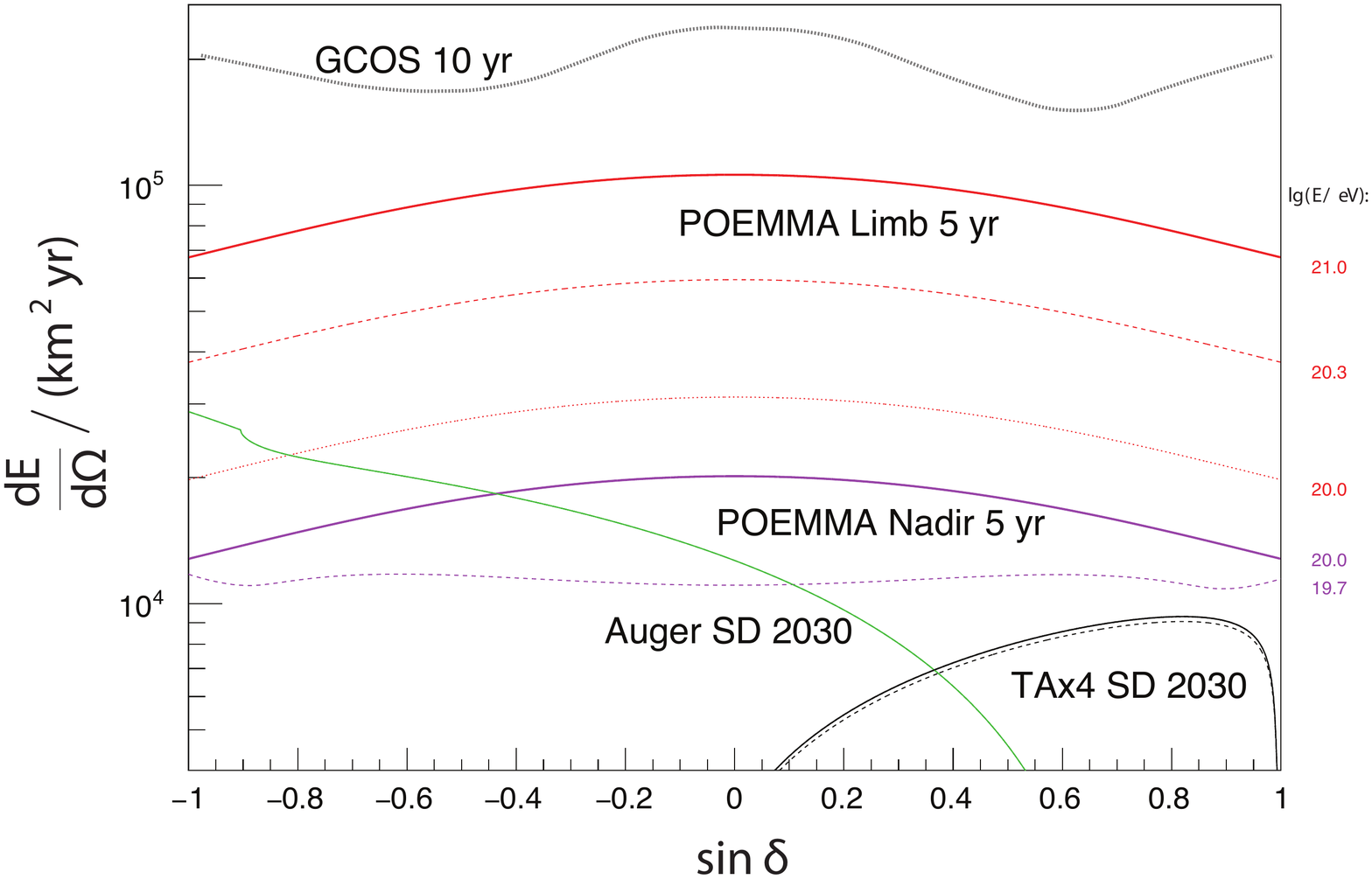}
  \vspace{-9mm}
  \Caption{Differential exposure as a function of declination, assuming a single-mode operation of POEMMA for the full 5-year benchmark. Purple curves denote the stereo (near-nadir) mode at $10^{19.7}$~eV (dashed) and $10^{20}$~eV (solid). Red curves denote the POEMMA
limb-viewing mode at $10^{20}$~eV (dashed), $10^{20.3}$~eV (dash), and $10^{21}$~eV (solid). The exposures of the surface detectors of the \pao and the Telescope Array (including the TA×4 upgrade) assuming being in operation until 2030 are shown as green and black curves, respectively. Also a rough estimate of a potential GCOS performance is indicated. Adapted from \cite{Anchordoqui:2019omw}.}
  \vspace{-3mm}
  \label{fig:space-ground}
\end{wrapfigure}

\paragraph{Complementarity of approaches}
Studying particles with space-based observatories \cite{gcos2021-poemma,Olinto:2020oky,Bertaina:2019zow} allows to cover enormous apertures with all-sky coverage. For example, the POEMMA concept foresees a fluorescence light telescope on a spacecraft in an orbit of 525~km. 
Looking downwards (nadir mode) for a 5~yr period will allow to reach exposures of $1-2 \times 10^4$~km$^2$\,yr. A second mode is mainly foreseen for neutrino detection, observing the Earth's limb with expected exposures up to $10^5$~km$^2$\,yr (with 5~yr operation) almost independent of declination. 
For GCOS a set of ground arrays with a total area of the order of 40\,000 km$^2$ is anticipated. Operating these arrays for a period of 10~yr would allow to reach exposures of the order of $2\times10^5$~km$^2$\,yr (for cosmic rays), as illustrated in \fref{fig:space-ground}.
The strength of space observations is the intrinsic full-sky coverage and the large aperture reached. A future ground-based observatory is expected to reach higher precision in terms of energy resolution and the identification of particle type ($\Xmax$).
Typical values for the resolution achieved in $\Xmax$ are of the order of 40~\gcm2 for spaced-based detectors \cite{Olinto:2020oky} compared to values better than 20~\gcm2 for ground based detectors.
Observing the same air volume (above GCOS) by POEMMA and, thus, observing the same particles simultaneously would allow for a cross calibration between the complementary techniques. 

An important aspect to be addressed for a future ground array is the assessment and optimization of the question ``large area vs high precision?''. With limited resources at hand, one needs to choose between a less dense array with a larger total area or a slightly smaller array with a denser detector spacing. A denser detector spacing will allow to achieve better values for the energy and mass resolution at the cost of less aperture, i.e., requiring longer exposure times to collect a sufficient number of particles at the highest energies.

This is directly connected to the question about the {\bf optimal/target energy range}.
Will the focus of GCOS be the fall-off region of the spectrum (above $10^{19.6}$~eV) or slightly lower energies above $10^{19}$~eV? Due to the steeply falling spectrum at the highest-energies the choice of threshold (respectively the main target energy range) has a big impact on the required area on the ground.
To identify and study the sources, will it be necessary to isolate low-$Z$ particles at the highest energies and track them back to their sources, i.e., conduct astronomy? This raises the immediate question: do light particles (protons, neutrinos, gamma rays) exist at all above $10^{19}$~eV? Although recent studies leave room for a small fraction of light particles at the higest energies \cite{vanVliet:2019nse}, this needs to be evaluated further by the ongoing experiments, such as the  \pao and the Telescope Array (and their upgrades).
If the mass composition of UHE cosmic rays follows (roughly) a Peter's cycle (as indicated, e.g., by \cite{Aab:2016zth,PierreAuger:2020kuy}) one would expect intermediate masses (He to CNO, with moderate charge numbers $2\le Z\le 8$) at energies around $10^{19}$~eV. Important for the further considerations here is the expected development of our understanding of Galactic and extra-galactic magnetic fields in the next decade. With sufficient knowledge of the magnetic fields it could be possible to correct for their effects for particles with moderate charge numbers ($Z\le 8$).
If this indeed will be possible after 2030 one could consider energies above $10^{19}$~eV as the main target of interest and would still be able to do particle astronomy, by back-tracing the particles in the Galactic  magnetic fields.
To converge on these issues we will closely follow the developments in the next years.

\paragraph{How to reach the physics case with a ground array?}
Identification of the UHE particle sources will require a good angular resolution.
A reasonable target is to achieve a resolution $<0.5^\circ$ at 100~EeV.
This will be determined by the grid spacing and the accuracy of the (GPS) timing at each station. Ionospheric distortions reduce the timing accuracy typically to values around $5-8$~ns. Assuming a detector spacing of the order of $1.6-2$~km an angular resolution $<0.5^\circ$ is realistic.

It is anticipated that GCOS will require good energy resolution of the order of $10-15\%$ to be able to investigate the energy spectrum in detail and discover new features/fine structure. In particular, in regions of a steeply falling energy spectrum, as e.g., at the highest energies a good energy resolution is important to restrict spill over of measured events to higher energies to an acceptable amount \cite{Brummel:2013urn}.
Good energy resolution is also important to identify and investigate transient sources.
A good target for GCOS could be to achieve 10\% at 100~EeV.
Experimentally, this will be determined by the spacing between detectors and the number of particles measured in each detector.

\begin{wrapfigure}{r}{0.7\textwidth} 
  \vspace{-5mm}
  \includegraphics[width=0.7\textwidth]{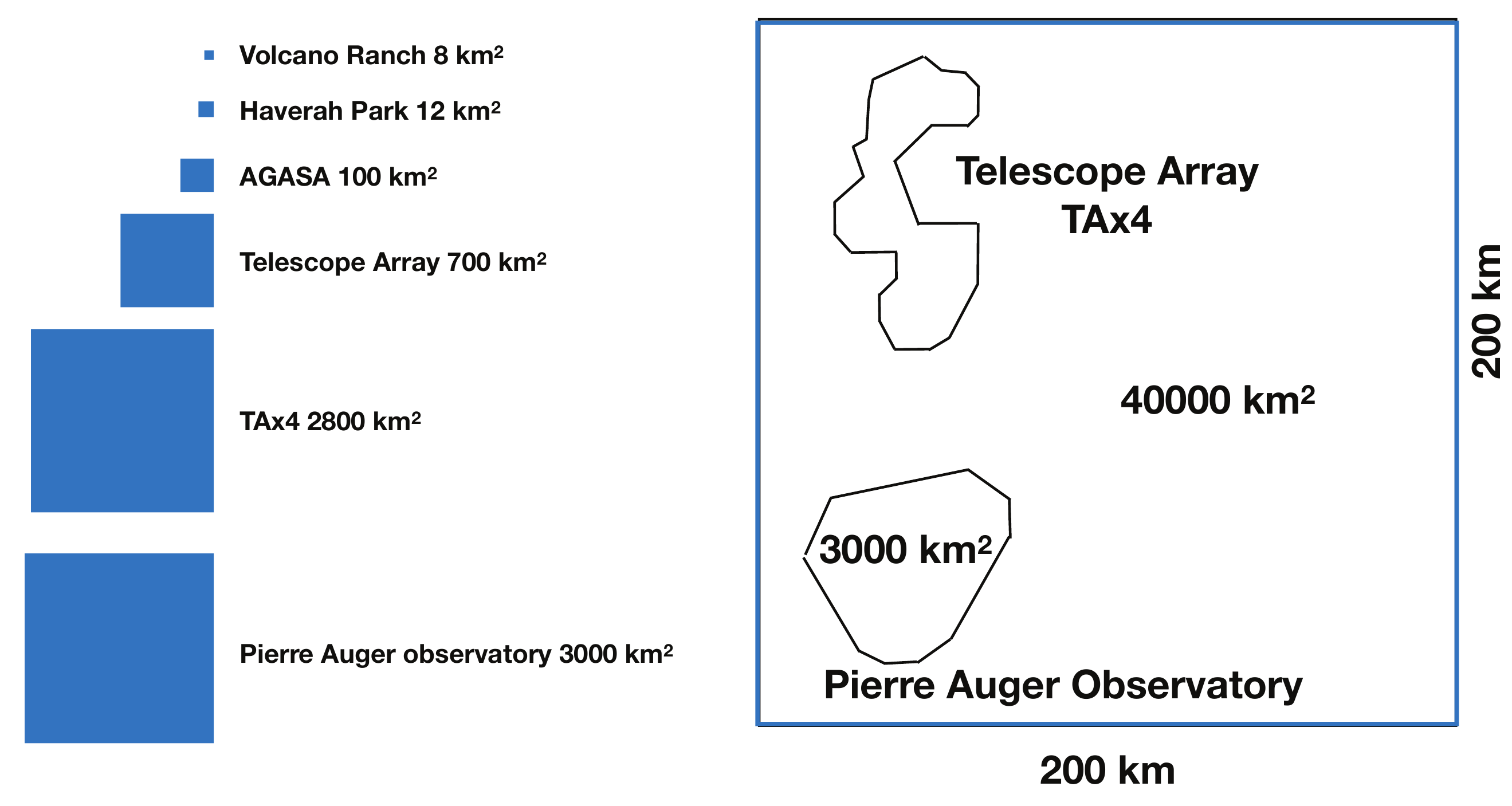}
  \vspace{-7mm}
  \Caption{Illustration of total collection area for existing installations and GCOS.}
  \vspace{-5mm}
  \label{fig:area}
\end{wrapfigure}

Another important aspect will be the capability to identify the particle type/atomic mass of each measured UHE particle.
This requires a good measurement of $\Xmax$ or the ratio of the electromagnetic to muonic particles in an air shower. Both quantities depend only logarithmicaly on the atomic mass of the primary particle. Assuming a Heitler-Matthews model, it can be shown that an uncertainty of $\Delta\ln A\approx1$ requires a quality in the measurement of $\Delta\Xmax\approx36$~\gcm2 or $\Delta(N_e/N_\mu)\approx16\%$ \cite{Hoerandel:2007fz}. This implies for an optimal case values around $\Delta\ln A\approx0.8$, i.e., 5 mass groups can be identified.
Experimentally, this is determined by the quality of the separation between the electromagnetic and muonic components  and the quality of the hadronic interaction model used to interpret the data.
Ultimately, GCOS will need to have excellent rigidity resolution. Since $R=E/Z$, this will require simultaneously good energy resolution of the order of $\sim10\%-15\%$ and good mass resolution with $\Delta\ln A <0.8$.
The charge $Z$ can only be derived indirectly, assuming $Z\approx A/2$.
This provides the foundation to find and study sources, but also to do particle and fundamental physics at extreme energies.

To collect a reasonable number of particles above a certain energy threshold within an acceptable time span will require a huge exposure.
A good target number for the total area of a ground array could be to aim for an order of magnitude larger area as the existing facilities, i.e., aim for about $40\,000-50\,000$~km$^2$. Realistic detector spacings could range from $1.6-2$~km, requiring about $10\,000-20\,000$ detection units. This is mainly driven by the costs per unit and constraint by the required resolutions. An important aspect is also the feasibility to build, deploy, operate, and maintain the units. Based on the experience with the reliable operation of the Telescope Array and the \pao an increase of the number of detection units by an order of magnitude beyond existing facilities seems to be feasible.

\paragraph{Complementarity of techniques}
To achieve good energy and mass resolution for UHE cosmic rays, GCOS will most likely be designed as hybrid detector, combining several detection concepts.
In parallel, other concepts are being developed as, e.g., a large array of radio antennas with the main objective to detect UHE neutrinos (GRAND) \cite{gcos2021-grand,Alvarez-Muniz:2018bhp}. Also this project foresees to have multiple sites. To maximize the synergies between the different projects it will be important to work together and unite and align the world-wide efforts to detect particles at UHEs and, thus, to maximize the exploration of the extreme Universe in the decades to come.

\section{Detection concepts}
Different detection concepts are at hand. They need to be optimized to reach the targeted physics case.
Fluorescence detectors  provide a calorimetric measurement of the shower energy and a direct and almost model-independent measurement of $\Xmax$. However, they have only a limited duty cycle ($\sim15\%$) due to constraints on atmospheric transparency and background light conditions.
An alternative with almost 100\% duty cycle is the use of radio antennas in a frequency range where the atmosphere is transparent to radio waves. Such detectors require radio-quiet regions.
The classical approach of a particle detector ground array has no restrictions with respect to radio interference or background light and the particle type is inferred from the ratio of secondary particles on the ground. Unfortunately, the conversion from measured signal ratios to the mass of the incoming particle requires Monte Carlo simulations and the result depends relatively strong on the hadronic interaction model used.

In the following a few detection concepts are being discussed. They serve as a starting point towards the development of a detailed plan and a realistic sketch of an anticipated concrete layout for GCOS.

\subsection{Advanced water Cherenkov detectors}
\begin{wrapfigure}{r}{0.55\textwidth} 
  \vspace{-5mm}
  \includegraphics[width=0.55\textwidth]{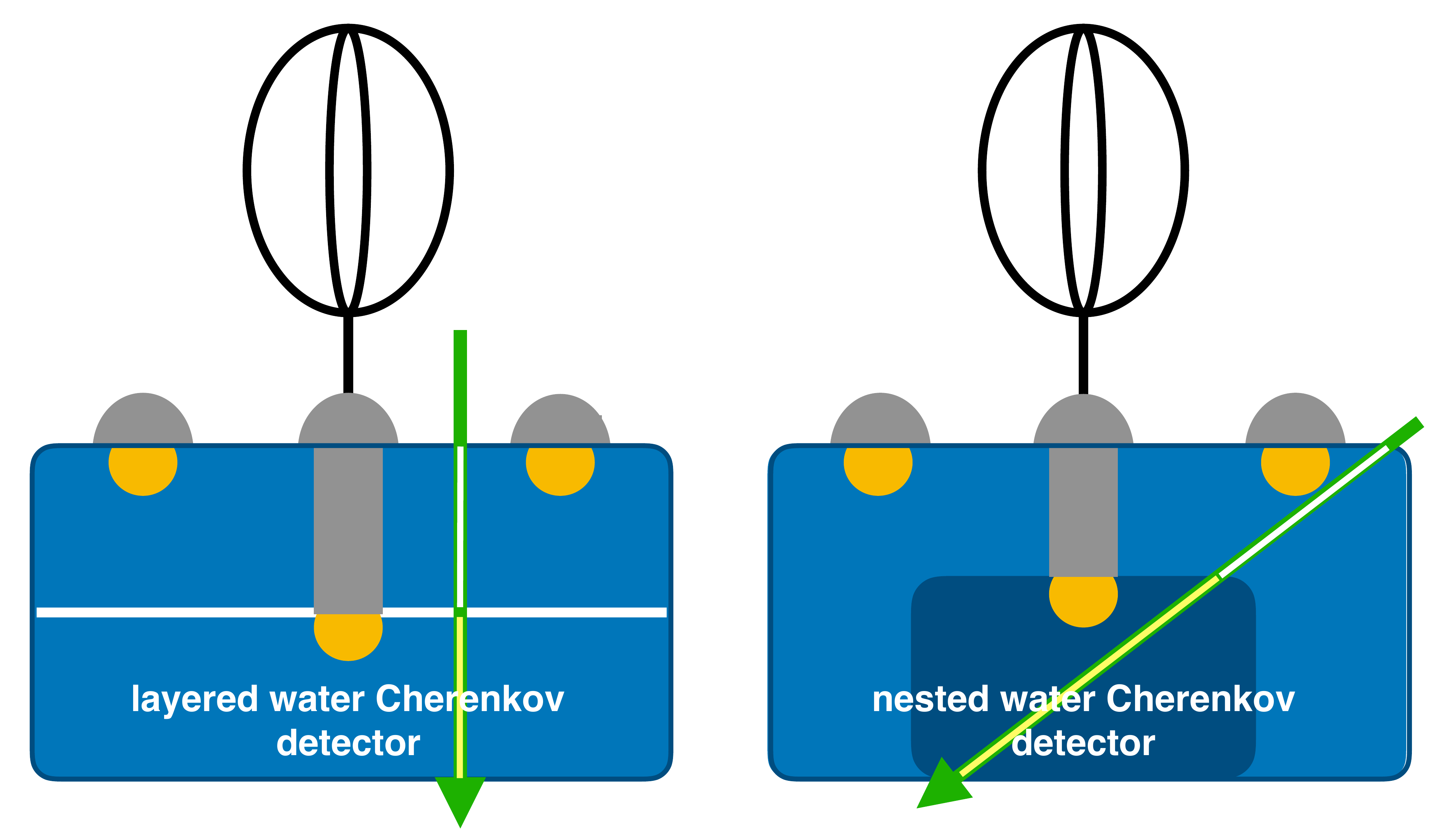}
  \vspace{-5mm}
  \Caption{Detection concepts, using a layered (left) and a nested (right) water Cherenkov detector with a radio antenna on top.}
  \vspace{-5mm}
  \label{fig:wcd}
\end{wrapfigure}
In order to determine the mass of each incoming cosmic ray  with a detector array one typically measures two shower components simultaneously, mostly the electromagnetic and muonic components are used. One can stack detectors on top of each other, as e.g., in the KASCADE experiment (two layers of scintillators with a lead-iron absorber in between) \cite{Antoni:2003gd} or in the AugerPrime upgrade \cite{Aab:2016vlz} (a layer of plastic scintillators on top of a water Chenerkov detector -- WCD).
The main idea is that the different shower components ($S_{\rm em}$ and $S_\mu$) generate different signals in the two sub detectors ($S_{\rm top}$ and $S_{\rm bot}$).
Using matrix inversion allows to derive $S_{\rm em}$ and $S_\mu$ from the measured values
$S_{\rm top}$ and $S_{\rm bot}$.
A cost effective approach is the use of layered water Cherenkov detectors \cite{Letessier-Selvon:2014sga,maris-segmented,gcos2021-wcd}. 
A big water volume is read out through optically separated segments as illustrated in \fref{fig:wcd}.
Prototypes of such detectors have been successfully operated at the \pao.

By carefully choosing the height-to-diameter ratio a WCD can be optimized to exhibit a more or less uniform detector response as a function of zenith angle, this is a big advantage over, e.g., a flat scintillator sheet.
If enhanced electron-muon separation is also desired for horizontal air showers (e.g.\ for neutrino detection) a possible design could be a nested detector (see \fref{fig:wcd}). The aspect ratio and the relative size of the inner and outer detector can be optimized to achieve a detector response with only a weak dependence on the zenith angle of the showers.
Layered or nested WCDs would be ideal for GCOS. They are very robust detectors, requiring not too much maintenance, as photosensors also SiPMs/APDs could be used, further reducing the maintenance efforts, they measure the mass composition with nearly 100\% duty cycle also in light-polluted regions, and they will allow the identification of gamma rays and neutrinos \cite{gcos2021-wcd}.

Significant progress is also expected from using machine-learning techniques to identify muons in WCDs \cite{Conceicao:2021xgn}.

\subsection{Radio detection}
Enormous progress has been achieved with the radio detection of air showers in the last decade \cite{Huege:2016veh,Schroder:2016hrv}. The technique is now considered mature and properties of cosmic rays are reconstructed from radio data \cite{gcos2021-radio}. Radio self-triggering works reliably in radio-quiet regions and providing a trigger from a particle detector allows the radio detection also in less radio-quiet areas. The radio emission physics is understood at the 10\% level. A very competitive measurement performance has been reached for the energy resolution ($10-15\%$), $\Xmax$ resolution ($15-20$~\gcm2), and the angular resolution ($<0.5^\circ$). In addition, the radio technique can be used to calibrate the absolute energy scale of a cosmic-ray detector.

Radio antennas on top of a particle detector, e.g., similar to the ones from the Auger Radio Detector \cite{Horandel:2019hrm,jrh-uhecr,Horandel:2019qwu,gcos2021-radio} are a very promising concept also for GCOS, see also \fref{fig:wcd}.
They provide a calorimetric measurement of the electromagnetic shower component with high precision ($\sim10\%$). In particular, this will allow to measure the electron-to-muon ratio for horizontal air showers in combination with a WCD.
Inclined air showers (with zenith angles above $60^\circ$) exhibit large radio footprints on the ground \cite{Aab:2018ytv}. An antenna array with a spacing of the order of $1.5-2$~km will allow for an effective measurement of showers with energies above $10^{18}$~eV. 
At the \pao the frequency range from $30-80$~MHz is investigated. Extending the band width to higher frequencies will allow single-station analyses, using also features like the slope of the frequency spectrum to derive the particle type.

Radio pulses also contain phase information, this allows to use interferometry, as successfully demonstrated by LOPES. A recent simulation study \cite{Schoorlemmer:2020low} suggests that interferometry can be used to derive $\Xmax$. Thus, in addition to measurements of the electromagnetic and muonic shower components, also $\Xmax$ would be available, ideal for tests of hadronic interaction models and particle physics applications (see above). The method necessitates outstanding time resolution ($\sim 1$~ns) in order to achieve a competitive resolution in $\Xmax$ \cite{Schluter:2021egm}. GPS timing needs to be corrected to achieve such accuracy.

\subsection{Fluorescence telescopes}
\begin{wrapfigure}{r}{0.35\textwidth} 
  \includegraphics[width=0.35\textwidth]{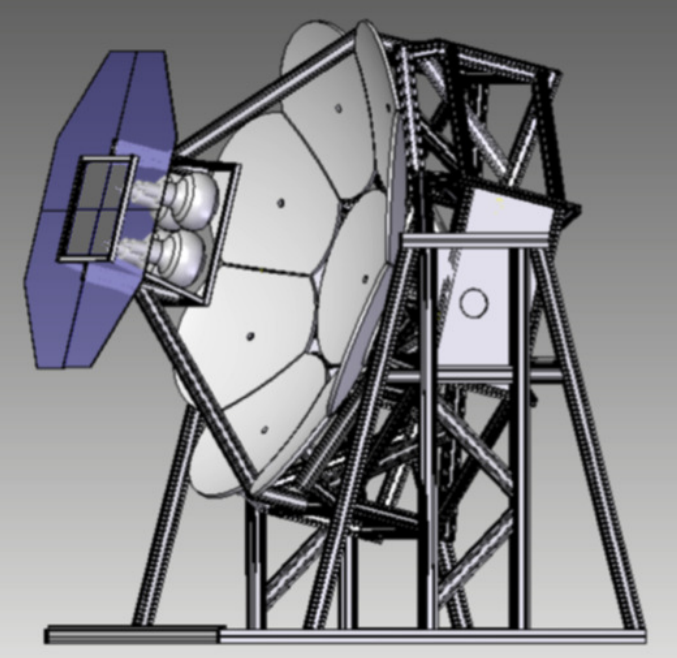}
  \vspace{-2mm}
  \Caption{A FAST telescope frame, showing four PMTs at the focus of a 1.6~m diameter segmented mirror. The support structure is made from aluminium profiles. The UV band-pass filter can be seen attached to the periphery of the camera box ~\cite{Malacari:2019uqw,Mandat:2017ura}.}
  \vspace{-3mm}
  \label{fig:fast}
\end{wrapfigure}

Fluorescence detectors could be included  to measure the calorimetric shower energy and the depth of the shower maximum, or, also a large stand-alone array of fluorescence detectors could be an option for a GCOS site.
The Fluorescence detector Array of Single-pixel Telescopes (FAST) is a design concept for a next-generation UHE observatory, addressing the requirements for a large-area, low-cost detector suitable for measuring the properties of UHE cosmic particles, having energies exceeding $10^{19.5}$~eV, with an unprecedented aperture \cite{Fujii:2015dra,Malacari:2019uqw,Mandat:2017ura}.
First prototypes have been developed, consisting of four 200~mm diameter photo-multiplier tubes at the focus of a segmented mirror of 1.6~m in diameter, shown schematically in \fref{fig:fast} and installed at the TA site in Utah. An automated all-sky monitoring camera has been developed to record cloud coverage and atmospheric transparency~\cite{Chytka:2020hgv}.

Over the last five years, the feasibility and reliability of the FAST model of fluorescence detection has been demonstrated, with the ultimate goal of laying the foundations for a future array with an order of magnitude larger ground coverage than previous-generation detectors targeted at the highest-energy cosmic rays. UHECRs with energies above $10^{19}$~eV have been measured and vertical laser signals to investigate the atmospheric transparency above the detector have been analyzed \cite{Malacari:2019uqw}. Further, a novel method for event reconstruction has been established that allows to circumvent the principal limitations of a coarsely-pixelized camera: the lack of timing information to tightly constrain the shower geometry. 
Continued operation will allow to further test the robustness of the FAST telescope concept, while work towards achieving full independence from the existing FD infrastructure will be continued, and in the process, FAST telescopes installed at both, the Telescope Array and \pao sites will allow to compare the quality of the atmosphere and sky between the two largest current-generation detectors.

\subsection{Further considerations}
\begin{wrapfigure}{r}{0.5\textwidth} 
  \includegraphics[width=0.49\textwidth]{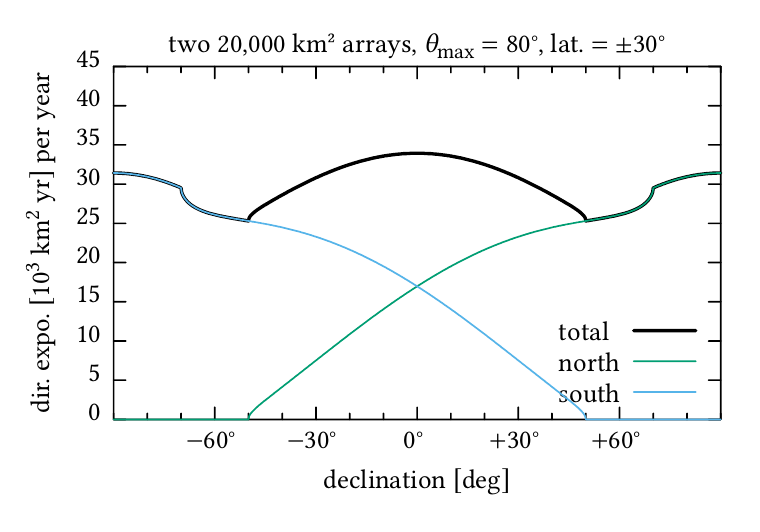}
  \vspace{-5mm}
  \Caption{Directional exposure for a pair of detectors, located at latitude$\pm30^\circ$ \cite{gcos2021-sites}.}
  \vspace{-5mm}
  \label{fig:exposure}
\end{wrapfigure}
\paragraph{Location of GCOS}
On of the most important decisions to be made will be the locations of the GCOS site(s).
Building an observatory with an exposure of the order of 40\,000~km$^2$ or more with a huge number of individual detector stations will most likely require to distribute the resources to build, maintain, and operate such an installation over several host countries/regions.
In order to achieve full-sky coverage from a single location, one needs to be on the equator and needs full $2\pi$ aperture. The celestial poles would only be detectable through horizontal air showers.
Thus, also from a scientific point of view it is useful to have several sites, located in the Northern and Southern Hemispheres in order to achieve an optimal sky coverage.
The optimal latitudes will depend on the number of sites to be implemented.
As illustration, choosing a pair of sites in the Northern and Southern Hemisphere, respectively, the optimum latitude 
is found to be $\pm30^\circ$ \cite{gcos2021-sites}.
This is illustrated in \fref{fig:exposure} for two sites with an area of 20\,000~km$^2$ each, covering zenith angles up to $80^\circ$.

Further constraints on the site locations arise from the detection principles used. For example, fluorescence telescopes will require clear atmospheres. A large ground array will require a region with low population density to ease negotiations with land owners, their number should be kept to a minimum. The implementation of radio detectors will require radio quiet regions.

\paragraph{Site logistics}
An important aspect of GCOS will be to investigate the properties of UHE particles on the full sky.
Thus, if GCOS will be composed of several sites, it will be crucial to implement the same detection concepts at different sites to make a cross calibration between the sites as simple as possible. It is also desirable that the same science groups are involved in the analysis of the data from different sites to avoid that different methods or assumptions lead to inconsistencies.

\paragraph{Societal impact}
For an observatory to be build after 2030 the societal impact will be of great importance.
We aim to implement GCOS as a ``green'' observatory, see e.g.\ \cite{Aujoux:2021ptv}.
Electrical power should be based on renewable resources, as, e.g., solar panels, as it is done already to a large extend at the Telescope Array and the \pao or wind energy.
GCOS sites will be located in regions with very low population density and most likely in economically weak regions.
The \pao increased the economic wealth of Malarg\"ue in Argentina and, in general, strengthened science in South America. 
In a similar way, we expect that the GCOS detector sites will have a positive impact on the development of the respective host regions.

\section{Next steps towards GCOS}
Based on the experience and scientific results which we have obtained and which we will obtain in the next decade, in particular with the \pao and the Telescope Array we will be in a position to make accurate estimates of what is required to build GCOS. In the immediate future, we will further develop the science case for multi-messenger astroparticle physics and will further specify the technical requirements for a next-generation observatory.

We plan to have a follow-up GCOS workshop around the end of 2021 with the aim of writing a roadmap towards a science case for UHE multi-messenger astroparticle physics beyond 2030. 
We also are discussing the option to have a {\sl workshop on the future of the field} attached to every upcoming meeting of the UHECR symposium series, as already done in 2018 \cite{uhecr2018-future}.

{\footnotesize
\setlength{\bibsep}{0.0pt}
\bibliographystyle{JHEP}
\bibliography{bibliography}
}

\clearpage
\section*{Full Author List: \Coll\ Collaboration}
{\parindent-3mm

\aaa{J\"org R. H\"orandel}{Radboud University}{Department of Astrophysics, IMAPP, P.O. Box 9010, 6500 GL Nijmegen}{Vrije Universiteit Brussel}{Brussel, Belgium}{Nikhef}{Amsterdam, The Netherlands}
\aaa{Rasha Abbasi}{Physics Department, Loyola University Chicago}{Chicago, IL, USA}{}{}{}{}
\aaa{Markus Ahlers}{Niels Bohr Institute}{Blegdamsvej 17, DK-2100 Copenhagen, Denmark}{}{}{}{}
\aaa{Kevin Almeida Cheminant}{Institute of Nuclear Physics Polish Academy of Sciences}{Radzikowskiego 152, 31-342 Krakow, Poland}{}{}{}{}
\aaa{Roberto Aloisio}{Gran Sasso Science Institute}{I-67100 L'Aquila}{INFN - Laboratori Nazionali Gran Sasso}{I-67010 Assergi (AQ), Italy}{}{}
\aaa{Rafael Alves Batista}{Radboud University}{Department of Astrophysics/IMAPP,  P.O. Box 9010, 6500 GL Nijmegen, The Netherlands}{}{}{}{}
\aaa{Neeraj Amin}{Karlsruhe Institute of Technology}{Institute for Astroparticle Physics, P.O. Box 3640, 76021 Karlsruhe, Germany}{}{}{}{}
\aaa{Gioacchino Alex Anastasi}{INFN Torino}{Via P. Giuria, 1, Torino, Italy}{}{}{}{}
\aaa{Luis Anchordoqui}{City University of New York}{The Graduate Center, 365 Fifth Avenue , New York, NY 10016, USA}{}{}{}{}
\aaa{Pedro Assis}{Instituto Superior T\'ecnico (IST); Laborat\'orio de Instrumenta\c{c}\~ao e F\'isica Experimental de Part\'iculas(LIP)}{Lisbon, Portugal}{}{}{}{}
\aaa{Aswathi Balagopal V.}{Wisconsin IceCube Particle Astrophysics Center}{University of Wisconsin, Madison, WI 53703, USA}{}{}{}{}
\aaa{Matteo Baittist}{University of Torino}{ Via P. Giuria, 10125 Torino, Italy}{INFN Torino}{Torino, Italy}{}{}
\aaa{Jose Bellido Cesares}{University of Adelaide}{Physics Department,  S.A. 5005, Australia}{}{}{}{}
\aaa{Douglas Bergman}{University of Utah}{Dept. of  Physics \& Astronomy, 115 S 1400 E, Salt Lake City, UT 84112, USA}{}{}{}{}
\aaa{Mario Edoardo Bertaina}{Department of Physics Univ. of Torino \& INFN Torino}{Via P. Giuria, 1 - 10125 Torino, Italy}{}{}{}{}
\aaa{Kathrin Bismark}{Karlsruhe Institute of Technology}{Institute for Atropraticle Physics, P.O. Box 3640, 76021 Karlsruhe, Germany}{}{}{}{}
\aaa{Pasquale Blasi}{Gran Sasso Science Institute}{Viale F. Crispi, 7, 67100 L’Aquila, Italy}{}{}{}{}
\aaa{Martina Boh\'a\v cov\'a}{Institute of Physics of the Czech Academy of Sciences}{Prague, Czech Republic}{}{}{}{}
\aaa{Denise Boncioli}{University of L'Aquila, Department of Physical and Chemical Sciences}{via Vetoio, L'Aquila, Italy}{INFN - LNGS}{Assergi (AQ), Italy}{}{}
\aaa{Mauricio Bustamante}{Niels Bohr International Academy, Niels Bohr Institute, University of Copenhagen}{Blegdamsvej 17, Copenhagen 2100, Denmark}{}{}{}{}
\aaa{Lorenzo Caccianiga}{Istituto Nazionale di Fisica Nucleare - Sezione di Milano}{Via Celoria 16 - Milano}{Universita degli studi di Milano, dipartimento di fisica "A. Pontremoli"}{Via Celoria 16 - Milano, Italy}{}{}
\aaa{Francesca Capel}{Technical University of Munich}{Boltzmannstr. 2, 85748, Garching, Germany}{}{}{}{}
\aaa{Rossella Caruso}{Universita di Catania}{Dipartimento di Fisica e Astronomia "Ettore Majorana", Via Santa Sofia, 64, 95123 Catania, Italy}{INFN Sezione di Catania}{Catania, Italy}{}{}
\aaa{Washington Carvalho Jr.}{Department of Astrophysics/IMAPP, Radboud University}{P.O. Box 9010, 6500 GL Nijmegen, The Netherlands}{}{}{}{}
\aaa{Antonella Castellina}{Osservatorio Astrofisico di Torino (INAF)}{Torino, Italy}{INFN, Sezione di Torino}{Torino, Italy}{}{}
\aaa{Gabriella Cataldi}{INFN Sezione di Lecce}{via per Arnesano, 73100 Lecce, Italy}{}{}{}{}
\aaa{Lorenzo Cazon}{Laboratorio de Instrumentacao e Fisica Experimental de Particulas (LIP)}{Lisbon, Portugal}{}{}{}{}
\aaa{Karel Cerny}{Palacky University in Olomouc, Faculty of Science, Joint Laboratory of Optics of Palacky University and Institute of Physics AS CR}{17. listopadu 12, 771 46 Olomouc, Czech Republic}{}{}{}{}
\aaa{Roberta Colalillo}{INFN, Sezione di Napoli}{Via Cintia, 80126 Napoli, Italy}{Università degli Studi di Napoli "Federico II"}{Dipartimento di Fisica ``E. Pancini'', Via Cintia, 80126 Napoli, Italy}{}{}

\aaa{Alan Coleman}{Bartol Research Institute and Dept. of Physics and Astronomy, University of Delaware}{Newark, DE 19716, USA}{}{}{}{}
\aaa{Ruben Conceicao}{Laboratorio de Instrumentacao e Fisica Experimental de Particulas (LIP)}{Lisbon, Portugal}{}{}{}{}
\aaa{Antonio Condorelli}{Gran Sasso Science Institute (GSSI), INFN}{Via Francesco Crispi 7, 67100, L'Aquila, Italy}{}{}{}{}
\aaa{Fabio Convenga}{Karlsruhe Institute of Technology}{Institute for Astroparticle Physics, P.O. Box 3640, 76021 Karlsruhe, Germany}{}{}{}{}
\aaa{Bruce Dawson}{University of Adelaide}{Physics Department,  S.A. 5005, Australia}{}{}{}{}
\aaa{Olivier Deligny}{Universit\'e Paris-Saclay}{Laboratoire de Physique des 2 Infinis Ir\`ene Joliot-Curie, CNRS/IN2P3,  Orsay, France}{}{}{}{}
\aaa{Sijbrand de Jong}{Radboud University}{Department of High-Energy Physics/IMAPP,  P.O. Box 9010, 6500 GL Nijmegen, The Netherlands}{Nikhef}{Amsterdam, The Netherlands}{}{}
\aaa{Ivan De Mitri}{Gran Sasso Science Institute (GSSI)}{Via Iacobucci 2, I-67100, L'Aquila}{INFN Laboratori Nazionali del Gran Sasso, Assergi, L'Aquila, Italy}{Via Acitelli 22, I-67100, Assergi, L'Aquila, Italy}{}{}
\aaa{Francesco de Palma}{Dipartimento di Matematica e Fisica "E. De Giorgi", Universita del Salento}{Lecce, Italy}{Istituto Nazionale di Fisica Nucleare, Sezione di Lecce}{I-73100 Lecce, Italy}{}{}
\aaa{Hans Dembinski}{TU Dortmund}{Experimental Physics 5a, Dortmund, Germany}{}{}{}{}
\aaa{Paolo Desiati}{University of Wisconsin}{WIPAC, Madison (WI), USA}{}{}{}{}
\aaa{Luca Deval}{Karlsruhe Institute of Technology}{Institute for Astroparticle Physics, P.O. Box 3640, 76021 Karlsruhe, Germany}{}{}{}{}
\aaa{Armando di Matteo}{Istituto Nazionale di Fisica Nucleare (INFN), Sezione di Torino}{Via Pietro Giuria 1, 10125 Torino, Italy}{}{}{}{}
\aaa{Rita de Cassia dos Anjos}{Universidade Federal do Paran\'a (UFPR)}{Departamento de Engenharias e Exatas, Pioneiro, 2153, 85950-000 Palotina, PR, Brazil}{Universidade Federal da Integra\c{c}\~ao Latino-Americana (UNILA)}{Programa de Mestrado em F\'isica Aplicada, Av. Silvio Am\'erico Sasdelli, 1842, 85867-670, Foz do Igua\c{c}u, PR, Brazil}{}{}
\aaa{Ralph Engel}{Karlsruhe Institute of Technology}{Institute for Astroparticle Physics, P.O. Box 3640, 76021 Karlsruhe, Germany}{Karlsruhe Institute of Technology}{Institute for Experimental Particle Physics, P.O. Box 3640, 76021 Karlsruhe, Germany}{}{}
\aaa{Johannes Eser}{University of Chicago}{Chicago, Illinois 60637, USA}{}{}{}{}
\aaa{Glennys Farrar}{New York University}{Center for Cosmology and Particle Physics, New York, NY, USA}{}{}{}{}
\aaa{Francesco Fenu}{University of Torino}{Universita degli Studi di Torino, Dipartimento di Fisica, Via Pietro GIuria 1, 10125, Torino, Italy}{}{}{}{}
\aaa{George Filippatos}{Colorado School of Mines}{1500 Illinois St, Golden, CO 80401, USA}{}{}{}{}
\aaa{Thomas Fitoussi}{Karlsruhe Institute of Technology}{Institute for Astroparticle Physics, P.O. Box 3640, 76021 Karlsruhe, Germany}{}{}{}{}
\aaa{Toshihiro Fujii}{Hakubi Center for Advanced Research, Kyoto University}{Sakyo-ku, Kyoto, Japan}{Graduate School of Science, Kyoto University}{Sakyo-ku, Kyoto, Japan}{}{}
\aaa{Christina Galea}{Radboud University}{Department of High-Energy Physics/IMAPP,  P.O. Box 9010, 6500 GL Nijmegen, The Netherlands}{}{}{}{}
\aaa{Maria Vittoria Garzelli}{Hamburg University}{II Institut f\"ur Theoretische Physik, Luruper Chaussee 149, 22761 Hamburg, Germany}{}{}{}{}
\aaa{Ugo Giaccari}{Department of Astrophysics/IMAPP, Radboud University}{P.O. Box 9010, 6500 GL Nijmegen, The Netherlands}{}{}{}{}
\aaa{Violeta Gika}{National Technical University of Athens}{9, Iroon Polytechnio 15780 Zografou  Greece}{}{}{}{}
\aaa{Christian Glaser}{Uppsala University}{Uppsala, Sweden}{}{}{}{}
\aaa{Noemie Globus}{Institute of Physics, Czech Academy of Sciences}{ELI Beamlines, 25241 Dolni Brezany, Czech Republic}{Flatiron Institute, Simons Foundation}{Center for Computational Astrophysics, New-York, NY10003, USA}{}{}
\aaa{Rob Halliday}{Michigan State University}{567 Wilson Road, East Lansing MI, 48824, USA}{}{}{}{}
\aaa{Balakrishnan Hariharan}{Tata Institute of Fundamental Research}{Homi Bhabha Road, Mumbai 400005, India}{The GRAPES-3 Experiment}{Cosmic Ray Laboratory, Raj Bhavan, Ooty 643001, India}{}{}
\aaa{Andreas Haungs}{Karlsruhe Institute of Technology}{Institute for Astroparticle Physics, P.O. Box 3640, 76021 Karlsruhe, Germany}{}{}{}{}
\aaa{Bohdan Hnatyk}{Taras Shevchenko National University of Kyiv}{Astronomical Observatory, 3 Observatorna str. Kyiv, 04053, Ukraine}{}{}{}{}
\aaa{Pavel Horvath}{Palacky University in Olomouc, Faculty of Science, Joint Laboratory of Optics of Palacky University and Institute of Physics AS CR}{17. listopadu 12, 771 46 Olomouc, Czech Republic}{}{}{}{}
\aaa{Miroslav Hrabovsky}{Palacky University in Olomouc, Faculty of Science, Joint Laboratory of Optics of Palacky University and Institute of Physics AS CR}{17. listopadu 12, 771 46 Olomouc, Czech Republic}{}{}{}{}
\aaa{Tim Huege}{Karlsruhe Institute of Technology}{Institute for Astroparticle Physics, P.O. Box 3640, 76021 Karlsruhe, Germany}{Vrije Universiteit Brussel}{Brussel, Belgium}{}{}
\aaa{Daisuke Ikeda}{Kanagawa University}{3-27-1 Rokkakubashi, Yokohama, Kanagawa, Japan 221-8686}{}{}{}{}
\aaa{P. Gina Isar}{Institute of Space Science}{Atomistilor 409, Magurele, Ilfov, 077125, Romania}{}{}{}{}
\aaa{Karl-Heinz Kampert}{University of Wuppertal}{Department of Physics, Gausstrasse 20, 42119 Wuppertal, Germany}{}{}{}{}
\aaa{Hanieh Karimi}{ Isfahan University of Technology}{Isfahan Province, Iran}{}{}{}{}
\aaa{Bianca Keilhauer}{Karlsruhe Institute of Technology KIT}{Institute for Astroparticle Physics, P.O. Box 3640, 76021 Karlsruhe, Germany}{}{}{}{}
\aaa{Abha Khakurdikar}{Radboud University}{Department of Astrophysics/IMAPP,  P.O. Box 9010, 6500 GL Nijmegen, The Netherlands}{}{}{}{}
\aaa{Eiji Kido}{Astrophysical Big Bang Laboratory, RIKEN}{Wako, Saitama, Japan}{}{}{}{}
\aaa{Jihyun Kim}{University of Utah}{High Energy Astrophysics Institute and Department of Physics and Astronomy,  Salt Lake City, UT 84112, USA}{}{}{}{}
\aaa{Varada Varma Kizakke Covilakam}{Instituto de Tecnologias en Deteccion y Astroparticulas (CNEA, CONICET, UNSAM)}{Centro Atomico Constituyentes, Comision Nacional de Energia Atomica, Av. General Paz 1499, San Martin, Buenos Aires, Argentina}{Karlsruhe Institute of Technology}{Institute for Astroparticle Physics, P.O. Box 3640, 76021 Karlsruhe, Germany}{}{}
\aaa{Matthias Kleifges}{Karlsruhe Institute of Technology}{Institut f\"ur Prozessdatenverarbeitung und Elektronik, P.O. Box 3640, 76021 Karlsruhe, Germany}{}{}{}{}
\aaa{Spencer R. Klein}{Lawrence Berkeley National Laboratory}{Berkeley CA 94720, USA}{University of California, Berkeley}{ Berkeley CA 94720, USA}{}{}
\aaa{Pavel Klimov}{Skobeltsyn Institute of Nuclear Physics, Lomonosov Moscow State University}{1(2), Leninskie gory, GSP-1, Moscow 119234, Russian Federation}{}{}{}{}
\aaa{Ramesh Koirala}{Nanjing University}{School of Astronomy and Space Science, Xianlin Road 163, Nanjing 210023, China}{Nanjing University}{11 Key laboratory of Modern Astronomy and Astrophysics, Ministry of Education, Nanjing 210023, China}{}{}
\aaa{Dmitriy Kostunin}{DESY}{Platanenallee 6, 15738 Zeuthen, Germany}{}{}{}{}
\aaa{Kumiko Kotera}{Institut d'Astrophysique de Paris}{Paris, France}{}{}{}{}
\aaa{Paras Koundal}{Karlsruhe Institute of Technology}{Institute for Astroparticle Physics, P.O. Box 3640, 76021 Karlsruhe, Germany}{}{}{}{}
\aaa{John Krizmanic}{NASA Goddard Space Flight Center}{Code 661 Greenbelt, Maryland 20771, USA}{University of Maryland}{Baltimore County, 1000 Hilltop Cir, Baltimore, MD 21250 USA}{Center for Research and Exploration in Space Science \& Technology}{USA}
\aaa{Vladimir Kulikovskiy}{INFN, Sezione di Genova}{Via Dodecaneso 33, Genova, 16146 Italy}{}{}{}{}
\aaa{Viktoria Kungel}{Colorado School of Mines}{Golden, CO, USA}{}{}{}{}
\aaa{Paolo Lipari}{INFN Roma Sapienza}{Roma, Italy}{}{}{}{}
\aaa{Francesco Longo}{University of Trieste}{Department of Physics, via Valerio 2, 34127 Trieste, Italy}{INFN sezione de Trieste}{Trieste, Italy}{}{}
\aaa{Lu Lu}{University of Wisconsin-Madison}{Madison, WI, USA}{}{}{}{}
\aaa{Volodymyr Marchenko}{Astronomical Observatory of the Jagiellonian University}{Orla 171, 30-244, Krakow, Poland}{}{}{}{}
\aaa{Federico Maria Mariani}{Universita degli Studi di Milano and INFN}{Dipartimento di Fisica,  Milano, Italy}{}{}{}{}
\aaa{Analisa Mariazzi}{Instituto de Física La Plata}{CONICET-UNLP, La Plata, Argentina}{}{}{}{}
\aaa{Ioana Maris}{Universite Libre de Bruxelles}{Bruxelles, Belgium}{}{}{}{}
\aaa{Giovanni Marsella}{Università degli Studi di Palermo}{Dipartimento di Fisica e Chimica "E. Segrè", via delle Scienze, 90128 Palermo, Italy}{INFN sez. Catania}{Catania, Italy}{}{}
\aaa{Eric Mayotte}{University of Wuppertal}{Department of Physics, Gausstrasse 20, 42119 Wuppertal, Germany}{}{}{}{}
\aaa{Daniela Mockler}{Universite libre de Bruxelles}{Boulevard du Triomphe 2, 1050 Bruxelles, Belgium}{}{}{}{}
\aaa{Pedro Henrique Morais}{Federal University of Paraiba}{Departamento  de  Fisica,  Universidade  Federal  da  Paraiba, Caixa  Postal  5008,  58059-900,  Joao  Pessoa,  PB,  Brazil}{}{}{}{}
\aaa{Ana Laura M\"uller}{ELI Beamlines, Institute of Physics, Czech Academy of Sciences}{25241 Dolni Brezany, Czech Republic}{}{}{}{}
\aaa{Katharine Mulrey}{Vrije Universiteit Brussel}{Brussel, Belgium}{}{}{}{}
\aaa{Kohta Murase}{Penn State University}{Department of Astronomy and Astrophysics, University Park, PA, USA}{}{}{}{}
\aaa{Marco Muzio}{Center for Cosmology and Particle Physics, Department of Physics, New York University}{726 Broadway, New York, New York, USA}{}{}{}{}
\aaa{Lukas Nellen}{I de Ciencias Nucleares, UNAM}{Circ. Ext S/N, 04510 Cd. Universidad, CDMX, Mexico}{}{}{}{}
\aaa{Anna Nelles}{Friedrich Alexander University}{Erlangen, Germany}{Deutsches Elektronen-Synchrotron DESY}{Zeuthen, Germany}{}{}
\aaa{Marcus Niechciol}{Universit\"at Siegen}{Naturwissenschaftlich-Technische Fakult\"at, Department Physik, 57068 Siegen, Germany}{}{}{}{}
\aaa{David Nitz}{Michigan Technological University}{Department of Physics, Houghton, MI 49931, USA}{}{}{}{}
\aaa{Angel Priyana Noel}{Astronomical Observatory of Jagiellonian University}{Orla 171, Krakow, Poland}{}{}{}{}
\aaa{Jesus Nunez}{Federal University of Lavras}{Lavras, Brazil}{}{}{}{}
\aaa{Shoichi Ogio}{Osaka City University}{Osaka, Japan}{}{}{}{}
\aaa{Foteini Oikonomou}{Norwegian University of Science and Technology}{Institutt for fysikk, Trondheim, Norway}{}{}{}{}
\aaa{Ek Narayan Paudel}{University of Delaware}{104 The Green, Office 217, Newark, DE, 19716, USA}{}{}{}{}
\aaa{Thomas Paul}{City University of New York}{Department of Physics and Astronomy, Lehman College, NY 10468, USA}{}{}{}{}
\aaa{Miroslav Pech}{Palacky University in Olomouc}{ Faculty of Science, Joint Laboratory of
Optics of Palacky University and Institute of Physics AS CR, 17. listopadu 12,
771 46 Olomouc, Czech Republic}{Institute of Physics of the Academy of Sciences of the Czech Republic}{Joint Laboratory of Optics of Palacky University and Institute of Physics AS CR,
17. listopadu 50a, 772 07 Olomouc, Czech Republic}{}{}
\aaa{Carmina Perez Bertolli}{ITeDA}{Buenos Aires, Argentina}{Karlsruhe Institute of Technology}{Institute for Astroparticle Physics, P.O. Box 3640, 76021 Karlsruhe, Germany}{}{}
\aaa{Lorenzo Perrone}{Universita del Salento}{Dipartimento di Matematica e Fisica "De Giorgi", Lecce, Italy}{INFN, Sezione di Lecce, via per Arnesano I-73100, Lecce, Italy}{}{}{}
\aaa{Tanguy Pierog}{Karlsruhe Institute of Technology}{Institute for Astroparticle Physics, P.O. Box 3640, 76021 Karlsruhe, Germany}{}{}{}{}
\aaa{Lech Piotrowski}{Faculty of Physics, University of Warsaw}{Pasteura 5, 02-093 Warsaw, Poland}{}{}{}{}
\aaa{Zbigniew Plebaniak}{Physics Department, University of Turin}{Via Pietro Giuria, 1, 10125 Torino TO, Italy}{INFN Turin}{}{}{}
\aaa{Bjarni Pont}{Department of Astrophysics/IMAPP, Radboud University}{P.O. Box 9010, NL-6500 GL Nijmegen, The Netherlands}{}{}{}{}
\aaa{Alessio Porcelli}{Universiteit Gent}{Departement Fysica en Sterrenkunde, Universiteit Gent, Proeftuinstraat 86 N3, 9000 Gent, Belgium}{}{}{}{}
\aaa{Julian Rautenberg}{University of Wuppertal}{Department of Physics, Gausstrasse 20, 42119 Wuppertal, Germany}{}{}{}{}
\aaa{Maximilian Reininghaus}{Karlsruhe Institute of Technology}{Institute for Astroparticle Physics, P.O. Box 3640, 76049 Karlsruhe, Germany}{Instituto de Tecnologias en Deteccion y Astroparticulas (ITeDA)}{Avenida General Paz 1499, San Martin, Provincia de Buenos Aires, Argentina}{}{}
\aaa{Marco Ricci}{INFN - Laboratori Nazionali di Frascati}{via E. Fermi 54, 00044 Frascati (RM), Italy}{}{}{}{}
\aaa{Frank Rieger}{Institute for Theoretical Physics, University of Heidelberg}{Philosophenweg 12, 69120 Heidelberg}{MPI for Nuclear Physics}{Heidelberg, Germany}{}{}
\aaa{Markus Risse}{University of Siegen}{Department of Physics, Walter-Flex-Str. 3, 57072 Siegen, Germany}{}{}{}{}
\aaa{Maria D Rodriguez Frias}{University of Alcala}{Space and Astroparticle Group, Madrid, Spain}{}{}{}{}
\aaa{Markus Roth}{Karlsruhe Institute of Technology}{Institute for Astroparticle Physics, P.O. Box 3640, 76021 Karlsruhe, Germany}{}{}{}{}
\aaa{Carsten Rott}{University of Utah}{Department of Physics and Astronomy, Salt Lake City, UT 84112, USA}{Sungkyunkwan University}{Deptartment of Physics, Suwon 16419, Korea}{}{}
\aaa{Alexandra Saftoiu}{"Horia Hulubei" National Institute for Physics and Nuclear Engineering}{Reactorului no 30, Magurele, Romania}{}{}{}{}
\aaa{Takashi Sako}{ICRR, University of Tokyo}{Kashiwanoha 5-1-5, Kashiwa, Chiba, Japan}{}{}{}{}
\aaa{Francesco Salamida}{Universita dell'Aquila, Dipartimento di Scienze Fisiche e Chimiche}{via Vetoio 10, 67100, L'Aquila, Italy}{INFN Laboratori Nazionali del Gran Sasso, Assergi (L'Aquila), Italy}{}{}{}
\aaa{Andrea Santanglo}{Eberhard Karls University}{Institut f\"ur Astronomie und Astrophysik, Kepler Center for Astro and Particle Physics, Sand 1, 72076 T\"ubingen, Germany}{}{}{}{}
\aaa{Eva Santos}{Institute of Physics of the Czech Academy of Sciences}{Na Slovance 1999/2, 182 21 Praha 8, Czech Republic}{}{}{}{}
\aaa{Fred Sarazin}{Colorado School of Mines}{Physics Department, Golden CO 80401, USA}{}{}{}{}
\aaa{Christoph M. Sch\"afer}{Karlsruhe Institute of Technology}{Institute for Astroparticle Physics, P.O. Box 3640, 76021 Karlsruhe, Germany}{}{}{}{}
\aaa{Harald Schieler}{Karlsruhe Institute of Technology}{Institute for Astroparticle Physics, P.O. Box 3640, 76021 Karlsruhe, Germany}{}{}{}{}
\aaa{Michael Schimp}{University of Wuppertal}{Department of Physics, Gausstrasse 20, 42119 Wuppertal, Germany}{}{}{}{}
\aaa{David Schmidt}{Karlsruhe Institute of Technology}{Institute for Astroparticle Physics, P.O. Box 3640, 76021 Karlsruhe, Germany}{}{}{}{}
\aaa{Frank Schroeder}{Bartol Research Institute, Departement of Physics and Astronomy, University of Delaware}{Newark DE, USA}{Institute for Astroparticle Physics, Karlsruhe Institute of Technology (KIT)}{Karlsruhe, Germany}{}{}
\aaa{Fabian Sch\"ussler}{IRFU, CEA, Universite Paris-Saclay}{F-91191 Gif-sur-Yvette, France}{}{}{}{}
\aaa{Valentina Scotti}{Dipartimento di Fisica - Universita degli Studi di Napoli Federico II}{Complesso Universitario di Monte Sant'Angelo - Via Cinthia, 21 - 80126 - Napoli, Italy}{INFN - Sezione di Napoli}{Napoli, Italy}{}{}
\aaa{Kenji Shinozaki}{National Centre for Nuclear Research}{28 Pulku Strelcow Kaniowskich 69, 90-559 Lodz, Poland}{}{}{}{}
\aaa{G\"unter Sigl}{Universit\"at Hamburg, II. Institut f\"ur theoretische Physik}{Luruper Chaussee 149, 22761 Hamburg, Germany}{}{}{}{}
\aaa{Dennis Soldin}{University of Delaware}{Newark, DE 19716, USA}{Bartol Research Institute}{Newark, DE 19716, USA}{}{}
\aaa{Glenn Spiczak}{University of Wisconsin River Falls}{River Falls, WI 54022, USA}{}{}{}{}
\aaa{Maximilian Stadelmaier}{Karlsruhe Institute of Technology}{Institute for Astroparticle Physics, P.O. Box 3640, 76021 Kalrsruhe, Germany}{}{}{}{}
\aaa{Rose Stanley}{Vrije Universiteit Brussel}{Pleinlaan 2, 1050 Brussel, Belgium}{}{}{}{}
\aaa{Yuichiro Tameda}{Osaka Electro-Communication University}{Neyagawa, Osaka, Japan}{}{}{}{}
\aaa{Caterina Trimarelli}{Universit\`a dell'Aquila, Dipartimento di Scienze Fisiche e Chimiche}{via Vetoio 10, 67100, L'Aquila, Italy}{INFN Laboratori Nazionali del Gran Sasso, Assergi (L'Aquila), Italy}{}{}{}
\aaa{Yoshiki Tsunesada}{Osaka City University}{3-3-138 Sugimoto, Sumiyoshi, Osaka, Japan 558-8585}{Nambu Yoichiro Instiute for Theoretical and Experimental Physics}{Japan}{}{}
\aaa{Ralph Ulrich}{Karlsruhe Institute of Technology}{Institute for Astroparticle Physics, P.O. Box 3640, 76021 Karlsruhe, Germany}{}{}{}{}
\aaa{Michael Unger}{Karlsruhe Institute of Technology}{Institute for Astroparticle Physics, P.O. Box 3640, 76021 Karlsruhe, Germany}{}{}{}{}
\aaa{Arjen van Vliet}{DESY}{Platanenallee 6, 15738 Zeuthen, Germany}{}{}{}{}
\aaa{Darko Veberic}{Karlsruhe Institute for Technology}{Institute for Astroparticle Physics, Postfach 3640, 76021 Karlsruhe, Germany}{}{}{}{}
\aaa{Peter Veres}{University of Alabama in Huntsville}{Huntsville, AL 35899, USA}{}{}{}{}
\aaa{Jakub V\'icha}{Institute of Physics of the Czech Academy of Sciences}{Prague, Czech Republic}{}{}{}{}
\aaa{Vadym Voitsekhovskyi}{Astronomical Observatory of Kyiv National University}{Kyiv, Volodymyrska st. 60, Ukraine}{}{}{}{}
\aaa{Serguei Vorobiov}{Center for Astrophysics and Cosmology, University of Nova Gorica,}{Vipavska 13, Nova Gorica, Slovenia}{}{}{}{}
\aaa{Yong-Gang WANG}{Shandong Agricultural University}{Daizongda Street No. 61, Tai'an, Shandong Province, China}{}{}{}{}
\aaa{Alan Watson}{University of Leeds}{Leeds, UK}{}{}{}{}
\aaa{Henryk Wilczynski}{Institute of Nuclear Physics, Polish Academy of Sciences}{ul. Radzikowskiego 152, 31-342 Krakow, Poland}{}{}{}{}
\aaa{Katsuya Yamazaki}{Chubu University}{1200 Matsumoto-cho, Kasugai, Aichi 487-8501, Japan}{}{}{}{}
\aaa{Alexey Yushkov}{Institute of Physics of the Czech Academy of Sciences}{Prague, Czech Republic}{}{}{}{}
\aaa{Orazio Zapparrata}{Universite Libre de Bruxelles}{Boulevard du Triomphe, 2 1050 Bruxelles, Belgium}{}{}{}{}
\aaa{Danilo Zavrtanik}{Center for Astrophysics and Cosmology, University of Nova Gorica,}{Vipavska 13, Nova Gorica, Slovenia}{}{}{}{}
\aaa{Mikhail Zotov}{Skobeltsyn Institute of Nuclear Physics, Lomonosov Moscow State University}{Moscow 119991, Russia}{}{}{}{}

}
%
%
%

\end{document}